\documentstyle[12pt]{article}

\catcode`\@=11
\long\def\@makefntext#1{
\protect\noindent \hbox to 3.2pt {\hskip-.9pt
$^{{\ninerm\@thefnmark}}$\hfil}#1\hfill}                

 \def\@makefnmark{\hbox to 0pt{$^{\@thefnmark}$\hss}}  

\def\ps@myheadings{\let\@mkboth\@gobbletwo
\def\@oddhead{\hbox{}
\rightmark\hfil\ninerm\thepage}
\def\@oddfoot{}\def\@evenhead{\ninerm\thepage\hfil
\leftmark\hbox{}}\def\@evenfoot{}
\def\sectionmark##1{}\def\subsectionmark##1{}}

\newcounter{sectionc}\newcounter{subsectionc}\newcounter{subsubsectionc}
\renewcommand{\section}[1] {\vspace{0.6cm}\addtocounter{sectionc}{1}
\setcounter{subsectionc}{0}\setcounter{subsubsectionc}{0}\noindent
	{\bf\thesectionc. #1}\par\vspace{0.4cm}}
\renewcommand{\subsection}[1] {\vspace{0.6cm}\addtocounter{subsectionc}{1}
	\setcounter{subsubsectionc}{0}\noindent
	{\it\thesectionc.\thesubsectionc. #1}\par\vspace{0.4cm}}
\renewcommand{\subsubsection}[1]
{\vspace{0.6cm}\addtocounter{subsubsectionc}{1}
	\noindent {\rm\thesectionc.\thesubsectionc.\thesubsubsectionc.
	#1}\par\vspace{0.4cm}}

\newcounter{appendixc}
\newcounter{subappendixc}[appendixc]
\newcounter{subsubappendixc}[subappendixc]

\renewcommand{\appendix}[1] {\vspace{0.6cm}
	\refstepcounter{appendixc}
	\setcounter{figure}{0}
	\setcounter{table}{0}
	\setcounter{equation}{0}
	\renewcommand{\thefigure}{\Alph{appendixc}.\arabic{figure}}
	\renewcommand{\thetable}{\Alph{appendixc}.\arabic{table}}
	\renewcommand{\theappendixc}{\Alph{appendixc}}
	\renewcommand{\theequation}{\Alph{appendixc}.\arabic{equation}}
	\noindent{\bf Appendix \theappendixc #1}\par\vspace{0.4cm}}

\def\abstracts#1{{
	\centering{\begin{minipage}{30pc}\tenrm\baselineskip=12pt\noindent
	\centerline{\tenrm ABSTRACT}\vspace{0.3cm}
	\parindent=0pt #1
	\end{minipage}}\par}}

\renewenvironment{thebibliography}[1]
	{\begin{list}{\arabic{enumi}.}
	{\usecounter{enumi}\setlength{\parsep}{0pt}
\setlength{\leftmargin 1.25cm}{\rightmargin 0pt}
	 \setlength{\itemsep}{0pt} \settowidth
	{\labelwidth}{#1.}\sloppy}}{\end{list}}

\newcounter{itemlistc}
\newcounter{romanlistc}
\newcounter{alphlistc}
\newcounter{arabiclistc}

\newcommand{\fcaption}[1]{
	\refstepcounter{figure}
	\setbox\@tempboxa = \hbox{\tenrm Fig.~\thefigure. #1}
	\ifdim \wd\@tempboxa > 6in
	   {\begin{center}
	\parbox{6in}{\tenrm\baselineskip=12pt Fig.~\thefigure. #1}
	    \end{center}}
	\else
	     {\begin{center}
	     {\tenrm Fig.~\thefigure. #1}
	      \end{center}}
	\fi}

\newcommand{\tcaption}[1]{
	\refstepcounter{table}
	\setbox\@tempboxa = \hbox{\tenrm Table~\thetable. #1}
	\ifdim \wd\@tempboxa > 6in
	   {\begin{center}
	\parbox{6in}{\tenrm\baselineskip=12pt Table~\thetable. #1}
	    \end{center}}
	\else
	     {\begin{center}
	     {\tenrm Table~\thetable. #1}
	      \end{center}}
	\fi}

\def\@citex[#1]#2{\if@filesw\immediate\write\@auxout
	{\string\citation{#2}}\fi
\def\@citea{}\@cite{\@for\@citeb:=#2\do
	{\@citea\def\@citea{,}\@ifundefined
	{b@\@citeb}{{\bf ?}\@warning
	{Citation `\@citeb' on page \thepage \space undefined}}
	{\csname b@\@citeb\endcsname}}}{#1}}

\newif\if@cghi
\def\cite{\@cghitrue\@ifnextchar [{\@tempswatrue
	\@citex}{\@tempswafalse\@citex[]}}
\def\citelow{\@cghifalse\@ifnextchar [{\@tempswatrue
	\@citex}{\@tempswafalse\@citex[]}}
\def\@cite#1#2{{$\null^{#1}$\if@tempswa\typeout
	{IJCGA warning: optional citation argument
	ignored: `#2'} \fi}}

\def\fnt#1#2{\footnotetext{\kern-.3em
	{$^{\mbox{\sevenrm #1}}$}{#2}}}

 1
 1
 1

\font\tenrm=cmr10
\font\tenit=cmti10

\font\ninerm=cmr9

\begin{document}

\begin{flushright} UCRHEP-T148\\July 1995\
\end{flushright}
\vspace{0.8cm}

\centerline{\bf CORRELATED NEUTRINO OSCILLATIONS}
\vspace{0.8cm}
\centerline{\tenrm ERNEST MA}
\baselineskip=13pt
\centerline{\tenit Department of Physics, University of California,}
\baselineskip=12pt
\centerline{\tenit Riverside, California 92521, USA}
\vspace{0.9cm}
\abstracts{In solar neutrino oscillations, if $\nu_e$ has a significant third
massive component, the allowed parameter space in $\Delta m^2$ and
$\sin^2 2 \theta$ for the first two components is shown to
be greatly increased.  This third component may be correlated to atmospheric
neutrino oscillations, as shown in a specific predictive seesaw model of
the $3 \times 3$ neutrino mass matrix.  Possible variations to include
the recent LSND results are briefly discussed.}
\vspace{0.8cm}
\rm\baselineskip=14pt
\section{Introduction}
There are now three categories of data which show evidence of neutrino
oscillations.  (a) In experiments which detect neutrinos from the sun,
there appears to be a deficit.  Hence $\nu_e$ has apparently disappeared.
(b) In experiments which measure the ratio $\nu_\mu/\nu_e$ in the
atmosphere, there also appears to be a deficit.  Hence a combination of
$\nu_e$ and $\nu_\mu$ diappearance and appearance may have also occurred.
(c) The recent results of the LSND (Liquid Scintillator Neutrino Detector)
experiment\cite{1} seem to indicate that $\nu_e$ has appeared where there is
originally only $\nu_\mu$.

Conventional interpretations of the above as neutrino oscillations always
plot $\Delta m^2$ versus $\sin^2 2 \theta$, assuming implicitly that only
two neutrinos are involved in each case.  As long as all mixing angles are
small, this is a good approximation because the $\Delta m^2$ in each case
is very different from one another.  However, the atmospheric data\cite{2}
are strongly indicative of a large mixing between $\nu_\mu$ and $\nu_e$ or
$\nu_\tau$ or both.  Hence $\nu_e$ may well be composed of three mass
eigenstates with a masive third component to account for all or part of
the atmospheric oscillations, whereas the solar oscillations are explained
by the first two components with a small $\Delta m^2$ together with
a nonnegligible contribution from the massive third component.  In the
following it will be shown that this has the important consequence of
enlarging the parameter space of $\Delta m^2$ and $\sin^2 2 \theta$ for
the first two components that is allowed by the present solar data.\cite{3}

Since the LSND results are indicative of a much larger $\Delta m^2$, of order
eV$^2$, a fourth neutrino is required if all of the data are to be explained
by neutrino oscillations.  This possibility will also be discussed.

\newpage
\section{Three-Neutrino Analysis of Solar Data}
The work that I will describe in this section was done in collaboration
with J. Pantaleone,\cite{4} who has been considering neutrino oscillations
of all three flavors for many years.\cite{5}  Other authors are now beginning
to follow suit.\cite{6}

Let the electron neutrino be a linear combination of three mass eigenstates:
\begin{equation}
\nu_e = \cos \theta ~\nu_1 - \sin \phi \sin \theta ~\nu_2 + \cos \phi
\sin \theta ~\nu_3,
\end{equation}
and assume\footnote{Note that if $m_3$ were of order a few eV, reactor data
would require that $\nu_3$ overlaps very little with $\nu_e$.  See the talk
by M. C. Gonzalez-Garcia, these proceedings.}
\begin{equation}
\Delta m^2_{13} \simeq \Delta m^2_{23} \sim 10^{-2}~{\rm eV}^2.
\end{equation}
Then for a given value of $\theta$, one may use the solar data to find the
allowed region in $\Delta m^2_{12}$ and $\sin^2 2 \theta_{e2} \equiv
4 |U_{e2}|^2 (1 - |U_{e2}|^2)$.  The results for $\sin^2 2 \theta = 0.35$
and 0.75 are shown below.
\vspace{10cm}

On the left is the allowed region for $\sin^2 2 \theta = 0.35$ which differs
from that of the two-neutrino analysis, {\it i.e.} $\theta = 0$, by only a
little.  However, there is already a firm indication that it
has enlarged.  On the right is the allowed region for $\sin^2 2 \theta = 0.75$
which shows dramatically that it has greatly increased and that the
adiabatic branch of the solution at around $\Delta m^2 = 10^{-4}~{\rm eV}^2$
is now allowed.  The dashed lines are theoretical predictions to be discussed
in the next section.

\newpage
\section{Seesaw Structure Revealed}
Recall that the well-known empirical relationship for the Cabbibo angle in
terms of the ratio of the $d$ and $s$ quarks, {\it i.e.} $\sin^2 \theta_C
\simeq m_d/m_s$, has led to the suggestion\cite{7} that
\begin{equation}
{\cal M}_{ds} = \left[ \begin{array} {c@{\quad}c} 0 & a \\ a & b \end{array}
\right].
\end{equation}
This simple observation has generated over the years an enormous literature
on quark mass matrices.  It is an especially active field of research in
the past two or three years.  Consider now a trivial extension of this
seesaw structure and apply it to the neutrino mass matrix, namely
\begin{equation}
{\cal M}_\nu = \left[ \begin{array} {c@{\quad}c@{\quad}c} 0 & 0 & 0 \\
0 & 0 & a \\ 0 & a & b \end{array} \right],
\end{equation}
but in the basis $\cos \theta ~\nu_e - \sin \theta ~\nu_\mu$, $\nu_\tau$,
and $\cos \theta ~\nu_\mu + \sin \theta ~\nu_e$.  For small $a/b$, the
mass eigenvalues are simply 0, $-a^2/b$, and $b$.  The usual three
neutrinos are related to the mass eigenstates by
\begin{equation}
\left( \begin{array} {c} \nu_e \\ \nu_\mu \\ \nu_\tau \end{array} \right) =
\left( \begin{array} {c@{\quad}c@{\quad}c} \cos \theta & -\sin \phi
\sin \theta & \cos \phi \sin \theta \\ -\sin \theta & -\sin \phi \cos \theta
& \cos \phi \cos \theta \\ 0 & \cos \phi & \sin \phi \end{array} \right)
\left( \begin{array} {c} \nu_1 \\ \nu_2 \\ \nu_3 \end{array} \right),
\end{equation}
where $\sin \phi \simeq a/b \simeq \sqrt {m_2/m_3}$.

The electron neutrino is then as given by Eq.~(1) and the discussion of
the previous section applies.  However, $\Delta m^2_{12}
= m_2^2$ is now correlated with $\sin^2 2 \theta_{e2}$ for a given
choice of $m_3$ which is of course constrained by atmospheric data.  In the
figures above, the dashed lines represent the predictions of Eq.~(4)
for $m_3 \simeq 0.17$ eV (left) and 0.063 eV (right).  They do indeed
intersect the allowed regions.

The atmospheric neutrino oscillations are given by
\begin{eqnarray}
P(\nu_\mu \rightarrow \nu_e) &=& {1 \over 2} \cos^4 \phi \sin^2 2 \theta
\left( 1 - \cos {{t \Delta m^2_{23}} \over {2p}} \right), \\
P(\nu_\mu \rightarrow \nu_\tau) &=& {1 \over 2} \sin^2 2 \phi \cos^2 \theta
\left( 1 - \cos {{t \Delta m^2_{23}} \over {2p}} \right).
\end{eqnarray}
Since the angle $\phi$ is small, $\nu_\mu$ oscillates mainly into $\nu_e$
in this simplest realization of the seesaw ansatz.  On the other hand,
the $\nu_\mu - \nu_\tau$ submatrix may be rotated without affecting
$\nu_e$, in which case a better fit to the atmospheric data can be obtained.

\section{A Specific Model}
To obtain Eq.~(4), start with $\nu_e$, $\nu_\mu$, $\nu_\tau$, and four
singlets: $\nu_S$, $N_1$, $N_2$, $N_3$.  Assume a discrete $Z_6$ symmetry
[$\omega^6 = 1$] and assign
\begin{equation}
(\nu_e, \nu_\mu, \nu_\tau) \sim (\omega, \omega^{-2}, 1); ~~~
(\nu_S, N_1, N_2, N_3) \sim (\omega, 1, \omega^2, \omega^{-2}).
\end{equation}
The Higgs sector is taken to consist of two doublets
$(\Phi_1, \Phi_2) \sim (1, \omega^{-3})$ and one singlet $\chi \sim \omega$.
The resulting $7 \times 7$ mass matrix is then given by
\begin{equation}
{\cal M}_7 = \left[ \begin{array} {c@{\quad}c@{\quad}c@{\quad}c@{\quad}c@
{\quad}c@{\quad}c} 0 & 0 & 0 & 0 & 0 & m_1 & 0 \\ 0 & 0 & 0 & 0 & 0 & m_2
& 0 \\ 0 & 0 & 0 & 0 & m_3 & 0 & 0 \\ 0 & 0 & 0 & 0 & m_4 & 0 & m_5 \\
0 & 0 & m_3 & m_4 & M_1 & 0 & 0 \\ m_1 & m_2 & 0 & 0 & 0 & 0 & M_2 \\
0 & 0 & 0 & m_5 & 0 & M_2 & 0 \end{array} \right],
\end{equation}
where $m_1$ comes from $\langle \phi_2^0 \rangle$, $m_{2,3}$ from
$\langle \phi_1^0 \rangle$, and $m_{4,5}$ from $\langle \chi \rangle$.
Large $M_{1,2}$ reduce the above to a $4 \times 4$ mass matrix
\begin{equation}
{\cal M}_4 = \left[ \begin{array} {c@{\quad}c@{\quad}c@{\quad}c} 0 & 0 & 0 &
m_1 m_5/M_2 \\ 0 & 0 & 0 & m_2 m_5/M_2 \\ 0 & 0 & m_3^2/M_1 & m_3 m_4/M_1 \\
m_1 m_5/M_2 & m_2 m_5/M_2 & m_3 m_4/M_1 & m_4^2/M_1 \end{array} \right].
\end{equation}
Assume now that $m_4^2/M_1$ dominates, then
\begin{eqnarray}
{\cal M}_3 &=& \left( \begin{array} {c@{\quad}c@{\quad}c} bs^2 & bsc & as \\
bsc & bc^2 & ac \\ as & ac & 0 \end{array} \right) \nonumber \\ &=&
\left( \begin{array} {c@{\quad}c@{\quad}c} c & 0 & s \\ -s & 0 & c \\
0 & 1 & 0 \end{array} \right) \left( \begin{array} {c@{\quad}c@{\quad}c}
0 & 0 & 0 \\ 0 & 0 & a \\ 0 & a & b \end{array} \right) \left( \begin{array}
{c@{\quad}c@{\quad}c} c & -s & 0 \\ 0 & 0 & 1 \\ s & c & 0 \end{array}
\right),
\end{eqnarray}
where $b = (m_1^2 + m_2^2)m_5^2 M_1/m_4^2 M_2^2$, $a = m_3 m_5 \sqrt {m_1^2
+ m_2^2}/m_4 M_2$, $s = m_1/\sqrt{m_1^2 + m_2^2}$, $c = m_2/\sqrt{m_1^2 +
m_2^2}$.  The desired seesaw structure is thus obtained.

\section{Addition of a Fourth Neutrino}
Since Eq.~(10) contains a fourth neutrino which couples to both $\nu_e$
and $\nu_\mu$, it may be considered as a candidate for explaining the
recent LSND results.\cite{1}  However, the required mass and mixing of this
singlet neutrino with $\nu_e$ are then too large to be consistent with
the nucleosynthesis bound on the number of light neutrinos.\cite{8}  To
avoid this problem, the most natural thing to do is to use the singlet
neutrino to explain the solar data in the matter-enhanced
small-angle nonadiabatic solution, as has been pointed out by many
authors.\cite{9}  In that case, $\nu_\mu$ and $\nu_\tau$ may be assumed
to have masses of a few eV, but a small enough mass difference and large
enough mixing to account for the atmospheric data.  A small mixing between
$\nu_e$ and $\nu_\mu$ may then be invoked to explain the LSND results.
A recently proposed model\cite{10} uses a discrete $Z_5$ symmetry and the
seesaw reduction of a $7 \times 7$ mass matrix to obtain four approximate
light neutrino mass eigenstates $\cos \theta ~\nu_e - \sin \theta ~\nu_S$,
$\cos \theta ~\nu_S + \sin \theta ~\nu_e$, $(\nu_\mu + \nu_\tau)/\sqrt 2$,
and $(\nu_\mu - \nu_\tau)/\sqrt 2$, with eigenvalues 0, $m_1$, $m_2$, and
$-m_2$ respectively.  In addition, mixing occurs between $\nu_e$ and
$\nu_\mu$, as well as between $\nu_S$ and $\nu_\tau$.  Note that $\nu_\mu$
and $\nu_\tau$ are pseudo-Dirac partners, hence $\sin^2 2 \theta = 1$ is
required for atmospheric neutrino oscillations.

To accommodate a fourth neutrino in the present context, a possible
variation is to double Eq.~(3) and consider the $4 \times 4$ mass matrix
\begin{equation}
{\cal M}'_\nu = \left[ \begin{array} {c@{\quad}c@{\quad}c@{\quad}c}
0 & a & 0 & 0 \\ a & b & 0 & 0 \\ 0 & 0 & 0 & c \\ 0 & 0 & c & d \end{array}
\right]
\end{equation}
in the basis $\cos \theta ~\nu_e - \sin \theta ~\nu_\mu$, $\nu_S$,
$\cos \theta ~\nu_\mu + \sin \theta ~\nu_e$, and $\nu_\tau$.
Solar neutrino oscillations are as given before, but now the second mass
eigenstate is mostly inert and there is no phenomenological constraint on
the ratio $a/b$
as in the case of Eq.~(4).  Atmospheric neutrino oscillations are mostly
between $\nu_\mu$ and $\nu_e$, whereas the LSND results are explained
by the fact that both $\nu_e$ and $\nu_\mu$ mix with $\nu_\tau$.  However,
because of the seesaw ansatz, the latter is correlated with
the former.  Numerically, they are indeed consistent with both sets of
data, although the value of $\Delta m^2$ in the LSND case is required to
be less than about 3 eV$^2$.

\section{Conclusions}
As more neutrino experiments accummulate more data, there are two important
messages for phenomenologists and model builders.  First, the naive assumption
that each case of neutrino oscillations is to be interpreted as between
only two mass eigenstates must be abandoned.  Atmospheric data tell us
that a large mixing angle exists between $\nu_\mu$ and $\nu_e$ or $\nu_\tau$
or both.  In this talk it has been shown that if $\nu_e$ has a significant
third massive component, the analysis of solar data allows a much larger
parameter space in $\Delta m^2$ and $\sin^2 2 \theta$ for the first two
components.

Second, the structure of the neutrino mass matrix is beginning to reveal
itself.  It is time to look for possible empirical relationships such as
the well-known $\sin^2 \theta_C \simeq m_d/m_s$ for quarks which may give
us a glimpse of the underlying theory of the origin of masses.
In this talk a first attempt, {\it i.e} Eqs.~(4) and (12), has been noted.

\section{Acknowledgements}
I thank Prof. J. W. F. Valle and the other organizers of the International
Workshop on Elementary Particle Physics: Present and Future (Valencia 95)
for their great hospitality and a
very stimulating program.  This work was supported in part by the U. S.
Department of Energy under Grant No. DE-FG03-94ER40837.

\end{document}